**Changes in mesoscale convective system precipitation structures in response to a warming climate**

Wenjun Cui[1,2], Thomas J. Galarneau, Jr.[2], and Kimberly A. Hoogewind[1,2]

1. *Cooperative Institute for Severe and High-Impact Weather Research and Operations, University of Oklahoma, Norman, OK*

2. *NOAA/OAR National Severe Storms Laboratory, Norman, OK*

Corresponding author: Wenjun Cui (wenjun.cui@noaa.gov)

**Abstract**

Mesoscale convective systems (MCSs) are crucial components of the hydrological cycle and often produce flash floods. Given their impact, it is important to understand how they will change under a warming climate. This study uses a satellite- and radar-based MCS tracking algorithm on convection-permitting climate model simulations and examines changes in MCS properties and precipitation structures between historical and future simulations. An underestimation in MCS total precipitation is evident in historical simulation compared to observations, due to model's depiction of MCS precipitation area and summertime occurrence frequency. Under pseudo-global warming, increases in MCS frequency and total warm season precipitation are observed, most notably in the southern U.S. The precipitation intensity and precipitating area generated by future MCSs also rises and results in an increase in precipitation volume. MCS precipitation structures are further classified into convective core and stratiform regions to understand how change in these structures contributes to future rainfall changes. In a warmer climate, the stratiform region demonstrates minimal change in size, but increases in mean precipitation rate and mean maximum precipitation rate by 15% and 29% are noted, respectively. A more robust future response is observed in the convective core region, with its size, mean precipitation rate and mean maximum precipitation rate increasing significantly by 24%, 37% and 42%, respectively. Finally, by examining the environmental properties of MCS initial condition, future intensification of convective rain may be attributed to a combined effect of substantial increases in atmospheric instability and moisture availability.

**Key Points**

- An MCS tracking algorithm is applied to regional climate models to examine the potential future changes in their rainfall characteristics

- An evaluation between observations and historical simulations shows that the model underestimates MCS total precipitation

- Future MCSs may become more frequent, more intense and produce higher rain volume, primarily due to the large changes in the convective core

**Plain Language Summary**

Thunderstorms that group together and grow to hundreds of kilometers in size and persist for more than a few hours are called mesoscale convective systems. They play an important role in Earth's water cycle and are often responsible for causing flash flooding events. Given their impact, understanding how they will change due to a warming climate is crucial. By analyzing the outputs from climate models, this study reveals future changes in various aspects of mesoscale storms behaviors, such as their frequency and where they occur. In addition, the rainfall patterns from the present and future mesoscale storms are compared. Our findings show that in a warmer climate, these storms will probably occur more frequently and bring more rain, especially in the southern parts of the United States. Individual mesoscale storms are also expected to contain larger rain areas and release heavier rainfall, which would contribute to an overall increase in rain volume. A shift from lighter rain to more moderate/heavy rain during these storms is also noticed, implying a potential rise in intense mesoscale storm events and heightened chances of flash floods in the future.

## 1. Introduction

Mesoscale convective systems (MCSs) are the largest of the convective cloud phenomena and develop from upscale growth and/or aggregation of cumulonimbus clouds (Houze 2004, 2019). In the central Great Plains of the United States, one of the most active regions for MCS activity in the mid-latitudes, these systems are responsible for producing approximately 30% to 70% of the total warm season precipitation (Cui et al. 2020; Fritsch et al. 1986; Haberlie and Ashley 2019; Nesbitt et al. 2006). Due to the large amount of precipitation and larger areal extent of precipitation that MCSs can produce, they are more conducive to heavy rainfall and flash flooding events compared to storms associated with isolated convection or non-convective stratiform rain (Hu et al. 2021). MCSs account for more than half of the floods during the warm season (Hu et al. 2020) and have caused some of the most fatal flash floods events in the central and eastern United States (Ashley and Ashley 2008). As a result, any changes in MCS properties due to climate change, such as changes in size, intensity, organization, and rainfall characteristics, can have severe consequences for the hydrological cycle, including the potential for increased flash flooding.

An increase in the frequency of intense and long-lived MCSs is observed over the central United States over the past four decades (Feng et al. 2016) and is aligned with the trend of warming springs. Moreover, convective storms may become more vigorous, producing higher rain rates in the future due to the increase in water vapor supply and additional latent heat release under a warming climate (Trenberth et al. 2003). This effect has been supported by future model projections (Hoogewind et al. 2017; Seeley and Romps 2014; Trapp et al. 2019; Prein et al. 2017b; Lepore et al. 2021).

Understanding the future changes in MCS precipitation properties such as intensity, amount, diurnal cycle, and spatial coverage largely depends on the model accuracy in representing the large-scale circulation and mesoscale convective processes associated with these convective systems. Substantial biases are found in traditional general circulation models (GCMs) in simulating MCSs. Due to their coarse resolution, GCMs are unable to resolve the phenomena at the convective scale and instead rely on convective parameterization schemes. Advances in computing resources and techniques have enabled modeling to be performed at convection-permitting ($\leq$ 4km) grid spacing without convective parameterization schemes, thereby improving the representation of MCSs in model simulations. Previous modeling studies suggest that high resolution (grid spacing $\leq$ 4km, temporal resolution ~1h) convection-permitting models (CPMs) can reproduce realistic MCS events compared to observations (Del Genio et al. 2012; Fan et al. 2017; Feng et al. 2018; Han et al. 2019; Zheng et al. 2023). Traditionally, CPMs driven by GCM outputs (i.e, the GCM provides initial and boundary conditions), also known as dynamical downscaling, are often used to examine how convective-scale weather phenomena responds to climate change (Ashley et al. 2023; Gensini et al. 2015; Gensini et al. 2022; Mahoney et al. 2013; Trapp et al. 2019; Hoogewind et al. 2017). An alternative approach, known as pseudo-global warming (PGW), directly imposes large-scale changes in the climate system on a control climate simulation by modifying the boundary conditions, and has been increasingly used in climate research (Brogli et al. 2023; Liu et al. 2017; Prein et al. 2017b; Rasmussen et al. 2011).

However, very limited work has been performed to study the future changes in MCSs based on convection-permitting climate models (CPCMs). Prein et al. (2017a) found that under the combined effects of increase in summertime MCS frequency, precipitating area, and precipitation intensity, the total precipitation volume associated with MCSs is projected to increase by 80% in the future under a Representative Concentration Pathway (RCP) 8.5 scenario over the central and eastern United States. Meanwhile over South America, Rehbein and Ambrizzi (2023) found that future MCSs may become less frequent but produce more precipitation in the Amazon basin.

MCSs are characterized by regions of convective and stratiform precipitation, and the variability of their structures has a significant impact on the size, amount, and intensity of the precipitation they produce. The spatial arrangement of convective and stratiform regions within MCSs determines their morphology and further influences their potential to produce severe weather events (Gallus et al. 2008; Parker and Johnson 2000, Pettet and Johnson 2003; Schumacher and Johnson 2005; Cui et al. 2021). Therefore, understanding the changes in convective and stratiform regions of MCSs is important to comprehend their impact on changes in MCS precipitation properties. In addition, the thermodynamic properties of convective and stratiform regions significantly differ from each other (Houze 2004). Changes in these properties could modify the heating profiles and result in changes in atmospheric circulation in a warmer climate.

The identification of convective and stratiform regions relies on the intensity of radar reflectivity ($Ze$; Feng et al. 2011; Starzec et al. 2017; Steiner et al. 1995). Prein et al. (2017a) and Rehbein and Ambrizzi (2023) identified MCSs based solely on hourly precipitation rate and brightness temperature without further classification into convective and stratiform regions. In a more recent work, Dougherty et al. (2023) studied the future changes in rainfall structure of eight flood producing MCSs from ensembles of historical and future simulations. They classified MCSs into deep convective cores, wide convective cores, and broad stratiform components and found that the future rainfall increase is mainly attributed to the increase in convective rainfall. Their study also reveals how changes vary between MCSs with different linear morphology. However, this study focuses on certain types of MCSs with a limited sample size and may not represent the wholesale response of MCSs to climate change.

The main objective of this study is to statistically analyze the changes in future MCS precipitation characteristics and understand how these changes are related to alterations in stratiform and convective regions within MCSs. To achieve this, this study will leverage the publicly available CPCM simulations produced by the National Center for Atmospheric Research (NCAR), along with a more accurate MCS tracking algorithm than those using single variables to track MCSs, to perform a more comprehensive investigation into the impacts of climate change on MCS properties for events occurring over a longer 9-year period. This study will also consider the impact of climate change on both MCS and non-MCS precipitation, a comparison not examined in previous research. This study will also analyze the changes in thermodynamic environments associated with MCS formation and further investigate the relationships between the changes in atmospheric conditions – such as moisture, instability, and shear – and the changes in MCS precipitation structures. This study primarily focuses on the thermodynamic effects associated with climate change on MCSs and does not account for potential changes in large-scale synoptic patterns due to the methodology employed in model simulations (see section 2.2). Wintertime MCSs are excluded from the study due to their relatively lower occurrence frequency and distinct meteorological environment conditions compared to MCSs during the warm season. The study period of 9-year may not be sufficient for a full representation of climate trends or decadal variability. Nevertheless, this study will address the gaps that remained unexplored in previous research. For instance, the tracking algorithm used in this study integrates both satellite and radar variables to examine both the cloud and internal components (stratiform and convective) of an MCS. This differs from Prein et al. (2017a) and Ambrizzi (2023), where the identification of MCS relied on a single variable. Additionally, the longer study period allows for a more comprehensive understanding of how MCS rainfall properties change statistically and climatologically, in contrast to the case study in Dougherty et al. (2023). The findings of this study will contribute to a better understanding of the complex interactions between climate change, MCSs, and the associated precipitation characteristics, and provide insights into the potential impacts of MCSs on the hydrological cycle and hazards in the future. This information may be important for policymakers, planners, and decision-makers in developing effective strategies for mitigating the impacts of extreme precipitation events – such as flash floods – and managing water resources in regions frequently affected by MCSs.

## 2. Datasets and Analysis Methods

This study focuses on MCSs occurring in March–October 2004–2012 and east of the Rocky Mountains within a domain bounded by 25° – 50°N and 110° – 75°W (land-only) for both observation and model analysis. These 9-years represent the overlapping period between observations and model. It should be noted that the analysis period exhibits a warmer and dryer climate than the longer term 1979–2023 climatology, particularly in the southern part of the study domain (not shown). This is likely associated with the occurrence of two strong and one moderate La Niña periods during the analysis period.

### 2.1 Observation datasets

Three observational datasets are used to identify observed MCS events: 1) The globally merged satellite IR brightness temperature ($T_b$) data produced by the National Center for Environmental Prediction's (NCEP)/Climate Prediction Center (Janowiak et al. 2017) available at 4-km × 4-km spatial resolution and half-hourly temporal resolution and covers the region between 60°S–60°N; 2) The 3-dimensional (3D) Next Generation Weather Radars (NEXRAD) radar reflectivity mosaic from GridRad (Bowman and Homeyer 2017) available for the contiguous United States (CONUS) with a spatial resolution of $0.02° × 0.02°$ and a vertical resolution of 1 km; 3) The Stage-IV multi-sensor hourly precipitation dataset produced by the River Forecast Centers (Lin and Mitchell 2005), which are computed from the Z-R relationship from over 150 NEXRAD sites and combines approximately 5500 hourly rain gauge measurements to produce precipitation analyses at 4-km resolution over the CONUS.

*2.2 Convection-permitting regional climate model simulations (CPCMs)*

High-resolution CPCM simulations (Rasmussen and Liu, 2017) produced by the NCAR Research Applications Lab with the Weather Research and Forecasting (WRF; Skamarock et al. 2008) model version 3.4.1 are used in this study. The model has a 4-km horizontal grid-spacing and 51 stretched vertical levels with a maximum altitude at 50 hPa. The output frequency varies depending on the variable, with 2-dimensional variables having an hourly resolution and 3D variables having a 3-hourly resolution. The model domain has $1360 × 1016$ grid points covering the majority of North America (Liu et al. 2017). The model contains a control simulation (CTRL) and a climate sensitivity simulation using the pseudo-global warming (PGW) approach. The CTRL simulation is forced by European Centre for Medium-Range Weather Forecasting Interim Reanalysis (ERA-Interim; Dee et al. 2011) every 6-hr for a continuous 13-year period from October 2000 to September 2013. Meanwhile, the PGW simulation is forced by ERA-Interim reanalysis plus a climate perturbation representing the RCP 8.5 scenario derived from a 19-model Coupled Model Intercomparison Project Phase 5 (CMIP5) ensemble monthly-mean climate change signal, which is applied to the same 13-year period:

$$WRF_{input} = ERA\text{-}Interim + \Delta CMIP5_{RCP8.5} \tag{1}$$

Here, $\Delta CMIP5_{RCP8.5}$ is the mean change signal of 95-year CMIP5 ensembles under the RCP8.5 scenario:

$$\Delta CMIP5_{RCP8.5} = CMIP5_{2071\text{-}2100} - CMIP5_{1976\text{-}2005} \tag{2}$$

The perturbed fields include horizontal wind, temperature, specific humidity, sea surface temperature, soil temperature, sea level pressure, and sea ice (Liu et al. 2017).

It should be noted that the RCP8.5 represents the "worst case" scenario that leads to a forcing level near the 90th percentile of baseline scenarios (van Vuuren et al. 2011). While it was not designated as an impossible outcome, it neither assumes a more nor less probable position than the other baseline scenarios, the majority of which yielded lower emissions (RCPs 2.6, 4.5, and 6). More information on the model design and configuration can be found in Liu et al. (2017).

## 2.3 MCS tracking

The FLEXible object TRacKeR (FLEXTRKR) algorithm developed by Feng et al. (2018) is used to track the MCSs from both observational datasets and CPCM simulations. This tracking algorithm first identifies and tracks large cold cloud systems (CCSs) with $T_b < 241$K (Maddox 1980), then further identifies MCSs based on the convective and precipitation features (PFs: contiguous areas with precipitation exceeding 1 mm hr$^{-1}$ and radar echoes at 2-km height > 17 dBZ) within the CCSs. A CCS will be labelled as an MCS when its area exceeds $6 \times 10^4$ km$^2$, containing a PF with major axis length > 100 km and convective feature with radar reflectivity > 45 dBZ at any vertical level, and having a lifetime longer than 6-hr. FLEXTRKR uses both cloud and precipitation features to identify MCSs, which is more rigorous and likely to produce more consistent and accurate results than the tracking methods that use cloud, precipitation or radar reflectivity features only (e.g. Huang et al. 2018, Stein et al. 2014, and Haberlie and Ashley, 2018a and 2018b).

FLEXTRKR classifies the tracked system into convective core (CC) and stratiform (SF) components based on the Feng et al. (2011) method. The composite Ze at each grid box is compared with a background intensity (Steiner et al. 1995). The background value is computed from the average over the central grid box and its surrounding grid boxes on all sides (4 km radius). A CC flag will be assigned to the grid box when it has a radar reflectivity greater than the background intensity and that exceeds 45 dBZ. Grid boxes that have echoes greater than 10 dBZ and are not classified as CC are labelled as SF.

To create a synthesized dataset for tracking MCSs from observations, the GridRad and Stage-IV datasets are interpolated to match the satellite IR $T_b$ 4-km grid. The detailed regridding process can be found in Feng et al. (2018). The observed MCS dataset (Atmospheric Radiation Measurement (ARM) user facility, 2004) will serve as a reference to assess the performance of model-simulated MCSs in CTRL run.

For performing the MCS tracking on model simulations, the total sky longwave upward flux at top (LWUPT) is first converted to $T_b$ based on the method developed in Yang and Slingo (2008). A quantile mapping transformation (Hyndman and Fan, 1996) is subsequently applied to the converted $T_b$ to match the distribution of observed $T_b$ to reduce the model bias in tracking CCSs. An example illustrating how $T_b$ distributions change before and after quantile mapping is applied can be found in Figure S1 in the online supplemental material. The same criteria are used for tracking MCSs here, except the composite radar reflectivity at 10 cm wavelength (REFLC_10CM) is used for PF identification as the model output does not provide 3-D radar reflectivity.

## 3. Results

*3.1 Model evaluation*

The simulated MCSs in the CTRL run are first compared with observed MCSs to evaluate whether their characteristics can be reasonably reproduced in model simulations. The mean MCS occurrence and days with MCS occurrence for the 2004–2012 period is shown in Figure 1 as a function of Julian day and month. In observations, a total of 2394 MCSs were identified during the study period, while in CTRL this number is reduced to 2263 (~-6%). The cumulative days with MCS occurrence (Fig. 1a) is similar in observation and CTRL at the end of October, with 195 days in the observations and 198 days in CTRL (Fig. 1a). This relatively small difference is a result of the offsetting between the overestimations in MCS days in spring and the underestimation in mid- to late- summer in CTRL (Fig. 1c). The CTRL run shows significantly higher cumulative MCS occurrence frequencies compared to observations for Julian days 90 to 210 (Fig. 1b). In the observations, the peak month of MCS occurrence day is July, with an average of 29 days MCS event days. MCS activity is the most frequent in June, with a mean monthly occurrence of 51 MCSs (Fig. 1d). However, CTRL shows an early peak in May for both the cumulative MCS occurrence days (30 days) and total activity (53 MCSs). When examining the monthly MCS occurrence number, the largest difference is seen in April, where the CTRL has 23 more MCSs occurring during this month than the observations. The model also significantly underestimated the MCS occurrence during July and August, with a negative difference of 18 and 17 events, respectively.

The spatial distributions of observed and simulated MCS and non-MCS total precipitation from March to August are shown in Figure 2. The spatial pattern of simulated MCS precipitation resembles the observed one, with higher precipitation amounts over the Central Plains and gradually decreasing to the surrounding regions (Fig. 2a). However, CTRL underestimated the MCS rainfall amount for the most parts of the study domain. This dry bias is the most severe in the Plains regions–Missouri, Kansas, Oklahoma, and Arkansas–where the highest precipitation amounts can reach around 550 mm in observations, but only half of that value in CTRL. This behavior is more clearly demonstrated in the scatterplot (Fig. 2c), where the model grid is interpolated to match the observational grid. A high pattern correlation coefficient of 0.905 is found between observation and model, and the model shows a consistent dry bias for precipitation amounts greater than 75 mm. For non-MCS precipitation, the model shows difficulties in estimating coastal and orographic precipitation, as the rainfall amounts are severely overestimated along the East Coast regions, Appalachians, and Rocky Mountains. The positive bias in mountainous regions can be partially attributed to errors in Stage-IV caused by radar beam blockage (Zhang et al. 2016). Despite this deficiency, CTRL can capture the distribution of non-MCS precipitating events in observations, and most paired grid points are along the 1-to-1 line in the scatterplot (Fig. 2f).

By breaking down MCS precipitation into different seasons (Fig. S1), the negative bias in CTRL in simulating MCS precipitation over the Central Plains is likely due to the underestimation in MCS events during mid to late summer as seen in Figure 1d. This is confirmed by comparing the spatial distributions of observed- and simulated-MCSs (not shown). This negative bias in MCS frequency is also found in Prein et al. (2017a) and Haberlie and Ashley (2018c), even though different tracking methods were used. Prein et al. (2017a) found that underestimations of MCS frequency typically occur under weak synoptic-scale forcing conditions, and this is potentially related to the poor representation of soil-atmosphere interaction caused by the warm temperature bias over the central US during summertime in the model found in Liu et al. (2017). Moreover, the precipitating area produced by MCSs is largely underestimated in CTRL (not shown) and leads to an overall underestimation in total precipitation over the study domain (Fig. 2c), although the warm season MCS frequency is similar in the observations and model. This could lead to an overall dry bias in regions with high MCS occurrence frequency, such as the southern US during spring (Fig. S2c) and southern Great Plains (SGP) in early fall (Fig. S2i), and further amplifying the dry bias during summer. This bias was not identified in Prein et al. (2017a) and is likely due to their MCSs being tracked using the method for object-based diagnostic (MODE) (Davis et al. 2006 and 2009) evaluation that incorporates the time dimension (MODE time domain or short MTD) (Clark et al. 2014). MTD tracks MCSs solely based on the hourly precipitation fields and employs a stricter rain rate threshold of 5 mm hr$^{-1}$, in contrast to this study's threshold of 1 mm hr$^{-1}$. Rain rates under 5 mm hr$^{-1}$ are often associated with stratiform rain, while underestimation of model-simulated stratiform rain area is found in most microphysical parameterizations (Feng et al. 2018; Han et al. 2019). The exclusion of light precipitation cells in MTD can result in underestimations of MCS precipitation area. Another difference between the tracking output from MTD and FLEXTRKR arises in the frequency of model-simulated MCSs during springtime. FLEXTRKR results in this study illustrate much higher frequency than the observations, whereas the MTD frequency is more comparable to observations. The discrepancy likely arises to the different choice in rain rate threshold. The contrast between spring (MAM) and summer (JJA) MCSs precipitation characteristics from Feng et al. (2019) and Song et al. (2019), wherein both studies MCSs are identified by FLEXTRKR, show that spring MCSs consist of larger stratiform areas but smaller convective areas than the summer ones. The precipitation rate probability distribution of spring MCSs is also less skewed compared to summer MCSs. Since the model already exhibits a dry bias in simulating MCS precipitation (Fig. 2c), the model-simulated springtime MCSs are even less intense than the observations. Consequently, these less intense cases are only detected by FLEXTRKR and not by MTD. Lastly, the bias in simulating MCS might arise from the model's inaccuracies in representing the physical and thermodynamic mechanisms linked to the formation of these systems. For instance, the underestimation observed along the Gulf of Mexico during spring (Fig. S2f) could result from a model bias in representing the low-level jet (LLJ) and transport of water vapor, which requires further investigation. As this study primarily focuses on the future changes in MCS precipitation properties, detailed evaluation of these formation mechanisms will not be extensively covered.

Despite the model deficiency in reproducing the MCS frequency and precipitating area, the physical processes such as convection initiation and upscale growth associated with MCSs are realistically captured, as suggested by the spatial pattern and diurnal cycle (Fig. S3) of precipitation. The main purpose of this study is to assess the impact of a thermodynamic warming climate on MCS precipitation structures from a statistical perspective as opposed to showing the capability of the model in representing the observed MCSs. These errors will be taken into consideration when discussing the results.

### 3.2 Changes in future MCS properties

The total MCS occurrence frequency increases by 6% in PGW compared to CTRL. A total number of 2543 MCSs are identified in PGW, relative to 2263 in CTRL. The left panel of Figure 3 shows the spatial distribution of MCS initial locations and density. Here, MCS initial time is defined as the first hour a CCS is detected in the $T_b$ of the tracked system. Compared to CTRL, the high MCS frequency region has shifted southward in PGW and become more concentrated over Kansas, Oklahoma, and northern Texas (Figs. 3a-c). More MCS events are also found along the Gulf and Southeast (SE) coasts. These differences are more clearly shown in Figure 3d, where positive difference in MCS frequency is found during most of the study period over the Southern

Great Plains (SGP) and SE. Except for the Northeast (NE), large annual variation in the change of MCS number is seen on a regional basis, and the change in total MCS number (black line) is mostly attributed to the SGP. For monthly differences (Fig. 3e), the MCS frequency has a substantial increase (>5) during midsummer in the SGP and SE and results in an increase in total number, indicating the MCS events become more active during summer in a warmer climate. An overall decrease in MCS frequency is found in the Northern Great Plains (NGP) and Midwest (MW) throughout the warm season. In the Northeast, the MCS frequency increases through late summer and decreases in early fall.

To examine whether climate change has differing impacts on MCS and non-MCS rainfall, changes in their fraction and total precipitation from CTRL to PGW are compared in Figure 4. Similar to Figure 2b, the contribution of MCS precipitation is the highest over the Great Plains, then gradually decreases toward the surrounding areas (Fig. 4a). In PGW, MCS contributes 4.7% more to the total precipitation than CTRL, with a significant increase in contribution over SGP and the western part of SE, and a decrease in parts of NGP and MW (Fig. 4b). The mean MCS precipitation amount increases by ~30% in PGW (Fig. 4c), which is more than the increased rate of MCS frequency (6%). This indicates that the increase in MCS total precipitation amount is more attributed to the changes in MCS cloud and precipitation properties than frequency. The spatial distribution showing significant changes in MCS precipitation amount corresponds well with those showing changes in MCS precipitation fraction. The mean non-MCS precipitation remains nearly the same in CTRL and PGW, with a small decrease of 0.33 mm in the future climate. The difference shows greater spatial variation than that of the MCS precipitation, exhibiting an overall significant decrease in the leeside of the Rocky Mountains that contributes to a significant increase in MCS precipitation fraction over southeastern New Mexico and west Texas, and a significant increase in the southern US. These findings suggest that the warmer climate has a larger impact on MCS than non-MCS events, and the southern U.S. may experience a significant increase in precipitation amount from both MCS and non-MCS in the future.

Figure 5 shows the averaged total precipitation produced by CTRL and PGW MCSs normalized by its centroid and separated into SF and CC regions. MCS precipitation exhibits a southwest to northeast tilt in shape, with slightly higher precipitation amounts being produced to the northeast of its center. The maximum rainfall amount and rainfall area increases in the future MCSs. More specifically, the averaged maximum precipitation from 96.3 mm in CTRL to 128.3 mm in PGW, a ~33% increase, respectively. For grids with precipitation rate greater than 5, 10, and 50 mm, the area increases by 15%, 18% and 56% from CTRL to PGW, respectively. The SF region contributes more to total precipitation volume than the CC region, due to its larger area and longer persistence compared to the CC region. The averaged maximum precipitation within the SF region increased by 15% from 62.7 mm to 72.2 mm, whereas the CC region, increased by 67%, from 26.5 mm to 44.2 mm. The positive shift in total precipitation produced by future MCSs for both SF and CC regions are statistically significant at the 95% confidence interval based on the Mann-Whitney U-test. These results suggest that the future MCSs will produce higher precipitation amounts and larger areas of precipitation, resulting in a substantially higher rainfall volume. The increase in total precipitation amount is attributed mostly to the large increase in CC precipitation. A more detailed analysis of changes in precipitation properties associated with SF and CC regions will be discussed in Section 3.3.

To determine the factors that may contribute to the increase in precipitation volume in future MCSs, we examine the changes in frequency of MCS lifespan, propagation speed, and maximum precipitation rate against the grid-point accumulated precipitation between CTRL and PGW. Grid-point accumulated precipitation refers to the average of the total number of rainfall points that belong to an MCS track. Lifespan is chosen as a factor because MCSs with longer lifetimes can produce more precipitation. In terms of propagation speed, an MCS with a higher speed can produce a larger precipitating area, while an MCS with slower movement will produce more localized precipitation. The maximum precipitation rate is directly related to the intensity of an MCS and can indirectly affect how much rainfall it produces. For reference, three selected variables from the CTRL run are validated against observation in Fig. S4. A weak linear relationship can be found between lifetime and grid-point accumulated precipitation. No obvious future change is seen in MCS lifetime as most have a lifetime between 12 to 24 hours in both CTRL and PGW (Figs. 6a-c). The frequency of MCSs with a grid-point accumulated rainfall less than 10 mm generally decreases, while the frequency of those with a value greater than 15 mm increases across most ranges of lifetime. This increase is statistically significant for rainfall ranges between 15 to 25 mm with a lifetime range from 12 to 24 hours. A weak negative linear relationship is observed between MCS propagation speed and grid-point accumulated precipitation and there is a trend towards slower movement and higher precipitation amounts in PGW, suggesting that future MCSs have a higher chance to produce localized precipitation and thus may be more conducive to flash flooding events (Figs. 6d-f). The observed increase in the frequency of slow-moving, high-intensity MCSs could be attributed to previously slow-moving MCSs producing a higher volume of rainfall in the future, since no noticeable change in the movement speed of MCSs has been observed in the PGW (not shown). Similarly, a decrease in the frequency of fast-moving, low-intensity MCSs should not be interpreted as a reduction in their actual frequency of occurrence. As expected, a strong linear relationship, with a correlation coefficient value ~0.7, is found between maximum rain rate and grid-point total precipitation (Figs. 6g-i). The frequency of paired values has shifted to higher values in PGW, with a noticeable decrease in frequency for MCSs with a precipitation rate below 50 mm h$^{-1}$ and a total precipitation below 15 mm. At the same time, an increase in frequency for MCSs with a precipitation rate between 75 and 125 mm h$^{-1}$ and a total precipitation between 15 and 25 mm is shown. This change may be even larger in the real atmosphere, as the maximum precipitation rates between 40 to 100 mm h$^{-1}$ are largely underestimated in model (Fig. S4c). In PGW, MCSs with a maximum precipitation rate over 200 mm h$^{-1}$ and a pixel total rainfall over 40 mm are present, which is not seen in CTRL. By analyzing the variables that could potentially contribute to changes in the precipitation volume of MCSs, we conclude that the primary factor is the change in precipitation intensity.

## 3.3 Changes in future MCS precipitation structures

In this section, the SF and CC regions of current and future MCSs are compared. CTRL MCSs have a mean SF area of 31467.2 km$^2$ and PGW has a mean value of 30292.6 km$^2$ (-4%; Fig. 7a). While there is an increase of maximum SF areal size in PGW, the distribution shows no obvious change and is not statistically significant at the 95% confidence interval according to the Mann-Whitney U test. The mean CC area in PGW increased from 4852.6 to 6016.4 km$^2$ (+24%) compared to CTRL, and the distribution shifts towards higher values, exhibits greater variation, and the change is significant at 95% confidence interval (Fig. 7c). To determine which life stage shows more significant changes in SF and CC areal size, their changes as a function of normalized lifetime are compared in Figs. 7b and d. CC areal size reaches its peak earlier than SF as the strong upscale growth of convection occurs during the early stages of MCS development, and SF is formed from the advection of ice particles from the CC and results in a delayed peak (Herzegh and Hobbs, 1980; Cui et al. 2019). Although no significant change is observed in the overall distribution, the decreases in SF size of future MCSs during the mature (stages 5 and 6) and dissipating stages (stages 9 and 10) are significant (Fig. 7b). Meanwhile, a significant increase of CC areal size is found throughout the MCS's lifecycle (Fig. 7d). The results here differ from Dougherty et al. (2023), where they found the average CC area increases by 75% and the average SF area increases by 40%. The difference can be attributed to the use of different convective and stratiform identification methods and differences in threshold values, as well as the differences in the model setup and physical parameterizations. Furthermore, their study only examined a few cases of MCSs with a certain linear structure, which may not be sufficient to represent climatological behavior. Since the simulations used in this study were performed over a larger domain and a longer period than Dougherty et al. (2023), after examining the model representation of these cases, we found some of the cases show noticeable differences between the observations and the CTRL run, and also between the CTRL and PGW runs. Therefore, it is difficult to compare the classification algorithms used in the two studies directly.

Figure 8 compares the probability distribution of precipitation rates for both MCS and non-MCS events. When examining all types of precipitation (left panel), MCSs generally consists of more heavy precipitation ($> 10$ mm h$^{-1}$) events than non-MCSs. A decrease in light precipitation ($< 2$ mm h$^{-1}$) and an increase in moderate and heavy precipitation rate are found in both MCSs and non-MCSs in PGW. The increase in precipitation rate $> 5$ mm h$^{-1}$ is more significant in MCS than non-MCS events. A similar change is observed in SF precipitation, except for an increase in precipitation less than 1 mm h$^{-1}$. With respect to CC precipitation, the frequency of light to moderate precipitation events decreases, while the frequency of heavy precipitation events increases. Again, these changes are more significant in MCS than in non-MCS precipitation, with the magnitude of the change almost three times greater in MCS precipitation. The MCS mean precipitation rate increases from 2.83 to 3.26 mm h$^{-1}$ (15.2%) for SF and from 8.92 to 12.24 mm

h$^{-1}$ (37.2%) for CC and are higher than the increase of 6.4% for SF and 14% for CC in non-MCS events. The findings here suggest that it is expected to have fewer weak to moderate MCSs and more intense MCSs in a warmer climate.

Because convective intensity is directly related to the radar reflectivity magnitude, the PDFs of CTRL and PGW composite Ze are compared (Fig. 9). Consistent with the changes in precipitation frequency, the lower reflectivity ranges occur less frequently in a future climate and higher reflectivity ranges are more frequent in both MCS and non-MCS events. However, the increase in composite Ze values skew to higher values in MCS than non-MCS events. For all MCS events, a decrease in the frequency of composite Ze values is observed in the 10−30 dBZ range is found. For non-MCS events, the decrease is between 0 and 25 dBZ. The greatest negative and positive changes occur around 22 (20) and 40 (25) dBZ in MCS (non-MCS) events, respectively. SF and CC regions contribute to a large increase in composite Ze around 40 dBZ and 60 dBZ, respectively, and the future warmer climate seems to have much less impact on the SF region in non-MCS than MCS events. For the CC region, a more pronounced increase in radar reflectivity between 50 to 65 dBZ is documented in MCS events, while the magnitude of change in frequency for high reflectivity greater than 65 dBZ is similar in MCS and non-MCS events. This similarity is likely due to the impact of global warming on convective events in general. These results are consistent with the findings of Rasmussen et al. (2020), where the same model simulations were used. Their results also suggest a decreasing frequency of weaker precipitating systems and increase in strong precipitating systems in a warmer climate.

The changes in maximum precipitation rate and composite Ze for SF and CC components are shown in Figure 10. The mean maximum SF precipitation rate for CTRL and PGW MCSs are 29.9 and 38.5 mm h$^{-1}$, respectively, representing an increase of 28.8%, and the change is significant at the 95% confidence level according to a Mann-Whitney U test. The SF precipitation rate has a wider range, and the upper limit increases from 152.4 to 214.9 mm h$^{-1}$. SF maximum composite Ze exhibits a somewhat different change pattern compared to precipitation rate. Although the upper limit has extended, more events consist of a lower maximum Ze in PGW. The mean value also decreases from 40.9 to 40.1 dBZ, which is different from the upward shifting of the entire SF Ze population (Fig. 9h). Meanwhile a noticeable future change in maximum CC precipitation rate is observed, with the entire distribution shifted to higher values. The mean value increases from 49.5 to 70.1 mm h$^{-1}$ (+41.6%) and the maximum value increases from 214.9 to 320 (+48.9%) mm h$^{-1}$. Similar changes are seen in the maximum Ze distribution, where the PGW exhibits a smaller variation of maximum Ze values compared to CTRL, and most events consist of a maximum Ze value greater than 70 dBZ. Mean maximum Ze increases from 68.6 to 72.0 dBZ from CTRL to PGW. For both scenarios, the maximum values are capped around 80 dBZ. The substantial increase in maximum CC precipitation rate and composite Ze values further suggests that intense MCS events will become stronger in a warming climate.

*3.4 Changes in future environments*

To understand the changes in thermodynamic properties that may contribute to the change in MCS convective rainfall structures, differences between CTRL and PGW most-unstable (calculated based on the entire vertical column) convective available potential energy (CAPE), most-unstable convective inhibition (CIN), precipitable water, and 0−3 km and 0−6 km vertical wind shear at 1−3 hours (the pressure level data is 3-hourly resolution) ahead of convective initiation for the different climate regions are shown in Figure 11. Each variable value is calculated based on the mean of an 80 km radius centered on the initial location of the MCS. The mean CAPE values (Fig. 11a) associated with MCS pre-initialization conditions have increased in the future, with the largest increase observed in the NE from 347.6 J kg$^{-1}$ to 749.3 J kg$^{-1}$ (+116%). The CAPE distribution has shifted toward higher values in the future and these positive changes are statistically significant at the 95% confidence interval based on a Kolmogorov-Smirnov (KS) test, except for NGP. The maximum CAPE value across the study domain increases from 4443.5 J kg$^{-1}$ in NGP to 5865.5 J kg$^{-1}$ (+32%) in Southeast from CTRL to PGW (outliers are not shown). CIN value exhibits an overall increase over the CONUS, primarily contributed by significant increases in NGP and MW regions. The increase in CIN may explain the future decrease in MCS frequency in these regions (Figure 3), as higher CIN values suppress convective initiation (CI). In addition, the maximum updraft intensity associated with MCS initiation is expected to increase in the future, as it can be approximated from CAPE value. Precipitable water also shows a notable increase as the atmosphere can hold more water vapor as it warms according to the Clausius-Clapeyron relation. Compared to CTRL, the mean value of precipitable water increased from 2.9 to 4.2 mm (+45%), and the 25$^{th}$ percentile value in PGW (2.6 mm) is nearly the same as the median value in CTRL (2.7 mm). Based on the KS test, these positive changes are significant at a 95% confidence interval in all regions, with more pronounced changes observed in the SE, NE, and MW. A slight decrease in vertical wind shear magnitude is found in both 0–3 (-1.3%) and 0–6 km (-7.7%) over the CONUS, which is expected due to a decrease in meridional temperature gradient (Trapp et al. 2019). However, the changes in 0−3 km shear exhibit some regional variations, with increases in shear magnitude observed in the NGP and MW, while decreases are seen in the remaining regions. Apart from the change in 0−6 km shear in SE, the changes in shear are not statistically significant at the 95% confidence interval. Additionally, changes in the spatial distribution of the LLJ and moisture availability, as indicated by the 850 hPa wind and mixing ratio, have been compared between the CTRL and PGW scenarios (not shown). Notably, moisture convergence has intensified over the SGP, and the LLJ has shifted more easterly in the future. This leads to increased interaction with the lee side of the Rocky Mountains, creating more favorable conditions for CI, and possibly contributes to the southward shift of MCS occurrence seen in Fig. 3c.

The seasonal variations (MAM, JJA, and SO) in the changes of selected variables are also investigated (not shown). Consistent changes in both CAPE and precipitable water across the different seasons are documented, although the magnitude of increase varies, with the most substantial increase occurring during JJA. There are variations in the change of vertical wind shear magnitude in the MW and NGP. In the MW, the 0–3 km shear exhibits a positive change during MAM and SO but turns negative during JJA. Meanwhile, in the NGP, the 0–3 km and 0–6 km shears are similar during MAM and JJA, but the difference becomes negative during SO, which could be affected by the limited sample size during SO.

The environments at MCS initial conditions are compared to the reference state climatology to further understand the anomalies (Fig. 12). The background climatology of selected environmental parameters is computed following the method presented in Nolan and McGauley (2012) and used in Galarneau et al. (2023) for tropical MCSs tracked by FLEXTRKR. To obtain the reference state of each MCS, 20 locations of other MCS initiations that took place on different dates and times within the study period are randomly selected. Next, the area-averaged means are calculated at each of these locations on the date and time of the reference MCS. The number of randomly selected locations is reduced from 40 to 20 in this study compared to Galarneau et al. (2023) because their study composites a larger study domain, specifically the entire tropics, whereas this study only focuses on MCSs occurring over the CONUS.

The climatological CAPE distribution (Fig. 12a) exhibits a similar pattern of change as seen in the MCS initial environments. It becomes more positively skewed with an increase in the upper range of CAPE values. Aside from NE region, the CAPE values associated with MCS initiations are higher compared to the climatological background state. The positive anomalies become more pronounced in PGW, particularly in MW and SE regions. While a future increase in CIN is seen in all regions (Fig. 12b), the CIN associated with MCS environments in SGP, SE, and NE remains almost unchanged in future and is not proportional to the change in background environment (Fig. 11b). Other factors, such as the increase in moisture availability, might also contribute to this effect. In another words, there may be a greater chance for MCSs to occur in extreme CAPE scenarios in the upper tail of the distribution. The similar CAPE value between MCS initiation and background conditions in the NE are likely due to the convective initiation being more affected by large-scale weather systems or elevated terrain than local thermodynamic perturbations. Increases in positive anomalies between the MCS initial and background climatology conditions are also evident in precipitable water (Fig. 12b). The increase in MCS precipitation amount corresponds well to the increase of precipitable water. As shown in Fig. 4c, the SGP, SE, and parts of the MW show significant increases in MCS precipitation amount, with the positive anomalies of precipitable water being more pronounced in these regions compared to others. Meanwhile no obvious changes are observed in 0–3 km (Fig. 12c) and 0–6 km (Fig. 12d) shear climatology distributions from CTRL to PGW. Compared to CTRL, the positive difference in shear between MCS initiation and background climatology becomes smaller in PGW. This change is probably related to the decrease in shear magnitudes in the Southern Great Plains and Southeast, which are the most favored locations for MCS events.

Based on the above results, it is noted that, except for CIN, the changes in MCS initiation environments are proportional to the changes in background climatology environments under climate warming. However, the increase in CAPE and precipitable water associated with MCS initiation is larger than the increase in the reference state, suggesting the future changes in thermodynamic environments likely favor the intensification of MCSs and therefore result in an enhancement of convective activity and precipitation generation within MCSs. The increase in the frequency of extreme precipitation events (Figs. 5, 8, and 10) associated with MCSs is likely linked to the statistically significant upward shift in the distribution of precipitable water in PGW, as extreme precipitation amounts are likely to scale positively and nonlinearly with higher values of precipitable water (Kunkel et al. 2020). The changes in shear environments are less noticeable compared to CAPE and precipitable water but are regionally dependent. This variability may be related to alterations in atmospheric flow during the initiation of MCSs and requires further investigation. Additionally, the morphology of an MCS is often determined by mid- to upper-level flow (Cui et al. 2019). Therefore, the slight changes in 0–6 km shear may indicate minor adjustments in MCS organization in the future. However, these assumptions require a more detailed analysis for conclusive insight.

## 4. Summary and discussion

In this study, climate warming effects on MCS properties and precipitation structure is investigated using high-resolution CPCM simulations. Warm season (March–October) MCSs from current and future (represented by change signal between the periods 2071–2100 and 1976–2005 under RCP 8.5 scenario) climates are identified and their associated properties are compared based on 9-year (2004–2012) statistical trends. The MCS precipitation is decomposed into convective and stratiform regions and analyzed separately. Additionally, the changes in pre-storm environments are investigated to assess their relationship to changes in MCS properties. Key findings are summarized below:

- Future MCS events could become more frequent and contribute more warm season precipitation compared to non-MCS events, with the changes being more noticeable over the southern and coastal regions. These changes are associated with a more robust response of thermodynamic environments to a warmer climate in these regions.

- The future MCSs had a larger precipitating area and higher precipitation rate, resulting in an increase in precipitation volume. The accumulated precipitation over similar temporal windows produced by future MCSs is also higher, making them more conducive to extreme rainfall and flash flooding events.

- CC and SF components exhibit different responses to climate change. The SF region shows slight decreases in size, but mean precipitation rate and mean maximum precipitation rate increase by 15% and 29%, respectively. Meanwhile, the size, mean precipitation rate and mean maximum precipitation rate of the CC region increase by 24%, 37% and 42% on average.

- The distribution of CC and SF precipitation rates shows a decrease in light to moderate precipitation events and an increase in heavy precipitation rate. This suggests that a warmer climate may be less conductive for weak to moderate MCS events, and more conductive for intense ones.

- The CC maximum precipitation rate is skewed with a longer tail of most intense precipitation in the future climate simulation, further indicating the most intense MCSs may become more intense in a warmer climate.

- By examining the pre-MCS environmental conditions, the intensification of CC precipitation is likely related to a combination of significant increases in atmospheric instability and moisture availability. The environments associated with MCS initial conditions show a more dramatic increase compared to the background climatology, suggesting the future environment favors the further enhancement of the most intense MCSs.

While this study is the first to investigate the changes in precipitation structures associated with MCSs from a climatological perspective, there exist several limitations to this research approach. First, there is a dry bias in simulating the total precipitation associated with MCSs over the central Plains. This is due to the underestimation of MCS frequency during midsummer, possibly resulting from the warm and dry biases and the poor representation of soil-atmosphere interaction in the model (Prein et al. 2017a). This bias may be partially accounted for (i.e., the bias is negligible as it exists in both CTRL and PGW simulations) since the study focuses on the changes between CTRL and PGW instead of between observations and PGW. The second is the underestimation in stratiform rain area, which is a known issue identified in previous studies (Feng et al. 2018; Dougherty et al. 2023) and is a common problem found within most microphysical parameter schemes (Fan et al. 2017). The presence of this model bias may result in misrepresentation of the future changes in SF regions and should be taken into consideration when interpreting the results of this study. Third, the PGW approach only considers the changes in CONUS-scale thermodynamic (i.e., temperature and humidity) and kinematic (i.e., wind) variables and does not account for potential changes in hemispheric-scale weather patterns such as changes in the position and intensity of subtropical highs and the frequency and tracks of extratropical cyclones. These factors could have significant impacts on the frequency, location, and intensity of MCSs in the future climate. Fourth, it is acknowledged that the 9-year study period is limited in representing the long-term climate trend and decadal variability. Fifth, despite using perturbations derived from 19 CMIP5 ensembles to generate the PGW simulation, it remains a single output.

This approach may introduce additional biases by not considering the model's internal variability. Finally, this work solely focuses on the "worst-case" RCP 8.5 scenario. Expanding the analysis to incorporate different RCP scenarios, such as RCP 2.6 and RCP 4.5, could offer valuable insights into the variability of results across various climate pathways, but is beyond the scope of this work. A linear relationship between the magnitude of the changes in MCS and emissions is probable when using the PGW approach, but it might not be the case when directly downscaling the future simulations. For instance, Haberlie et al.'s (2023) dynamical downscaling of the RCP 4.5 and RCP 8.5 scenarios from the Community Earth System Model showed different changes in regional annual MCS precipitation between the two scenarios.

At the same time, there are several insights that can be drawn from this work. First, building upon the findings in Dougherty et al. (2023), this study analyzes a larger quantity of MCS events from regional climate model simulations, offering a comprehensive view of how the precipitation structure of MCS is expected to evolve in a warmer climate. One of the major findings of this study is the identification of statistically significant changes in convective precipitation intensity and area produced by future MCSs. Secondly, through the comparison of climatological reference states with MCSs in this study, it is noted that thermodynamic environments, including instability and moisture availability, are projected to become more favorable for convective development and precipitation generation in the future than the current state. Importantly, these changes exhibit regional dependence, where the changes in southern parts of the U.S. experiencing more pronounced alterations compared to the other regions. Such insights underscore the potential for more intense and larger MCSs, which can have significant implications for water resources management and flood risk assessment in regions prone to MCS activity. Lastly, this study reveals how the use of different tracking inputs (i.e., satellite and radar products, and precipitation estimations) and the selection of threshold values can influence the climatological representation of MCSs in model simulations and subsequently, impact the analysis results. For example, in Prein et al. (2017a), the model's underestimation of stratiform area is not captured due to their use of a higher precipitation rate as a threshold. Furthermore, both their algorithm and the method used in Haberlie and Ashley (2018c) yield a lower frequency of MCS occurrence during spring than this study and can likely attributed to similar reasons as springtime MCSs tend to produce less intense precipitation rate.

Overall, the findings from this study provide valuable insight into potential future changes in MCS precipitation properties. The future changes in MCS population, location, and intensification associated with convective rainfall found in this study have implications for policymakers and urban planners, as they provide important information for adapting to and mitigating the potential impacts of climate change on extreme rainfall and flash flooding events. Since this study only focuses on the precipitation characteristics associated with MCSs and given that MCSs are often responsible for producing other types of weather hazards including hail, severe wind, and tornadoes, a natural extension of this study would be to investigate the potential changes in these hazards associated with MCSs in a warming climate.

**Acknowledgment**

The funding of this study is provided by NOAA/Office of Oceanic and Atmospheric Research under the NOAA-University of Oklahoma Cooperative Agreement #NA210AR4320204, U.S., Department of Commerce. We appreciate Dr. Adam Clark and four anonymous reviewers for their invaluable insights and feedback on this paper.

**Figure captions**

**Figure 1.** Upper panel (a) shows the accumulated days with MCS occurrence and (b) accumulated MCS occurrence as a function of Julian day from March to October. Lower panel (c) shows the days with MCS occurrence per month and (d) monthly MCS occurrence from March to October. Green line represents observed MCSs and blue line represents model simulated MCSs in the CTRL run. Shaded area represents the values within the 95% confidence interval.

**Figure 2.** Spatial distribution of warm season (March – October) MCS and non-MCS total precipitation from observations and CTRL run. The total precipitation is the accumulated precipitation from March through October from MCS or non-MCS. The right panel shows the scatter density plot between observed and model-simulated warm season total precipitation for MCS and non-MCS events. The Pearson pattern correlation coefficient (r) is also shown by the scatterplot.

**Figure 3.** Top panel (a) shows the climate regimes defined in this study. Mid-left panel shows the spatial distributions of MCS initial locations color-coded with probability density (scatter density) from (b) CTRL and (c) PGW simulations. Each dot represents the initial location associated with an MCS. Right panel shows the differences (PGW-CTRL) of (d) yearly (March – October) and (e) monthly MCS occurrences between PGW and CTRL runs for different regions: North Great Plains (NGP), Southern Great Plains (SGP), Southeast (SE), Midwest (MW), and Northeast (NE). The dots in the monthly occurrence difference plot represents the 95% confidence interval lower and upper interval values.

**Figure 4.** Spatial distributions of warm season (March–October) MCS precipitation fraction in (a) CTRL and (b) fraction difference between PGW and CTRL runs, and warm season total precipitation difference between PGW and CTRL runs for (c) MCS and (d) non-MCS events. Grey dots represent the grid values that exceed the two-tailed 95% confidence interval based on the bootstrapping method.

**Figure 5.** Normalized total precipitation (>= 1 mm) from an MCS through its lifetime relative to the storm center (0, 0) from CTRL (left panel) and PGW (right panel) simulations for all (top row), stratiform (mid row) and convective (bottom row) precipitation. All precipitation consists of both stratiform and convective precipitation, as well as all other unclassified precipitation.

**Figure 6.** Heatmap showing the frequency between (upper) MCS lifespan, (middle) propagation speed, and (lower) maximum precipitation rate against the pixel accumulated precipitation for CTRL, PGW, and their (PGW – CTRL) difference. R value represents the Pearson correlation coefficient between the two paired variables. Grey x mark represents the difference is significant at 95% confidence interval based on the bootstrapping method.

**Figure 7.** Left panel shows the violin plots of the averages of (a) stratiform areal (SFA) size, and (c) convective core areal (CCA) size for CTRL and PGW MCSs. Each dot represents the mean value of SFA or CCA associated with a model-simulated MCS. Right panel shows the (b) SFA and (d) CCA change as a function of MCS's composite lifecycle. The shaded area represents the 25$^{th}$ to 75$^{th}$ percentile range, the triangle indicates the life stage that the difference between PGW and CTRL is significant at 95% confidence interval based on a Mann-Whitney U test.

**Figure 8.** Probability density functions of MCS and non-MCS precipitation rate for CTRL and PGW runs, and the difference between two simulations. The comparison is performed for all, and stratiform (SF) and convective core (CC) components of precipitation events. All type consists of both SF and CC pixels, as well as all other precipitating pixels that are unclassified.

**Figure 9.** Same as in Figure 8, but for composite radar reflectivity.

**Figure 10.** Violin plots showing the distributions of maximum precipitation rate and composite radar reflectivity of each MCS in CTRL and PGW for stratiform (SF) and convective core (CC) regions. Each box within the violin shape shows the 25th–75th percentile with the median indicated by the black horizontal line.

**Figure 11.** Boxplot showing the distributions of area-averaged (a) MUCAPE, (b) MUCIN, (c) preciptiation water, (d) 0-3 km bulk vertical wind shear, and (e) 0-6 km bulk vertical wind shear for MCS pre-initial condition for different climate regions. The left and right boxes of the same color correspond to the values associated with CTRL and PGW MCSs, respectively, for the respective climate region. The box extends from the 25$^{th}$ to the 75$^{th}$ percentile of the data, and the whiskers extend 1.5 times the inter-quartile range from the box. The black line represents the median value, and the black dotted line represents the mean value. The black dot above the x-axis label name represents the selected variable at specified region, the change from CTRL to PGW is significant at 95% confidence interval based on the Kolmogorov-Smirnov test.

**Figure 12.** Same as in Fig. 11 but for background climatology without the significance test. Triangle and square symbols represents the median and mean of the MCS pre-initial environments in Fig.11, repetively.

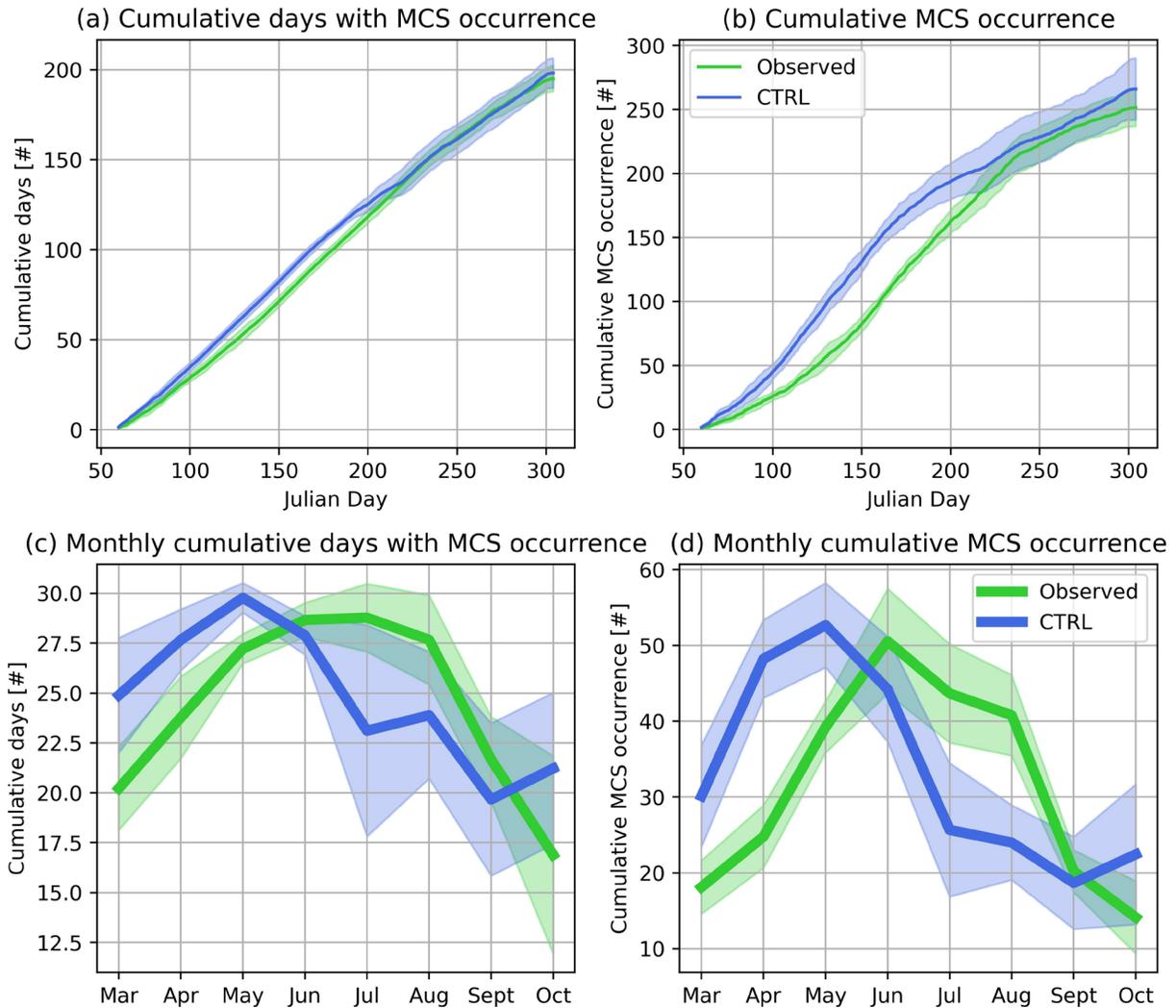

**Figure 1.** Upper panel (a) shows the accumulated days with MCS occurrence and (b) accumulated MCS occurrence as a function of Julian day from March to October. Lower panel (c) shows the days with MCS occurrence per month and (d) monthly MCS occurrence from March to October. Green line represents observed MCSs and blue line represents model simulated MCSs in the CTRL run. Shaded area represents the values within the 95% confidence interval.

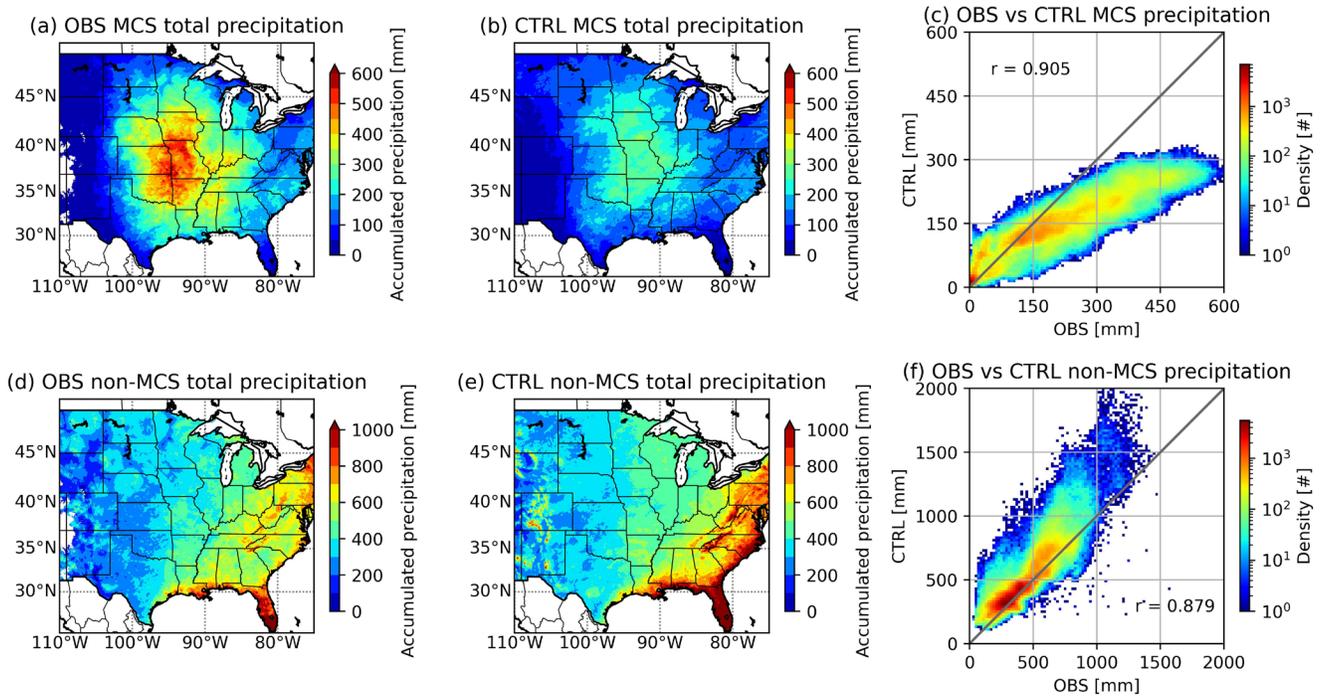

**Figure 2.** Spatial distribution of warm season (March – October) MCS and non-MCS total precipitation from observations and CTRL run. The total precipitation is the accumulated precipitation from March through October from MCS or non-MCS. The right panel shows the scatter density plot between observed and model-simulated warm season total precipitation for MCS and non-MCS events. The Pearson pattern correlation coefficient (r) is also shown by the scatterplot.

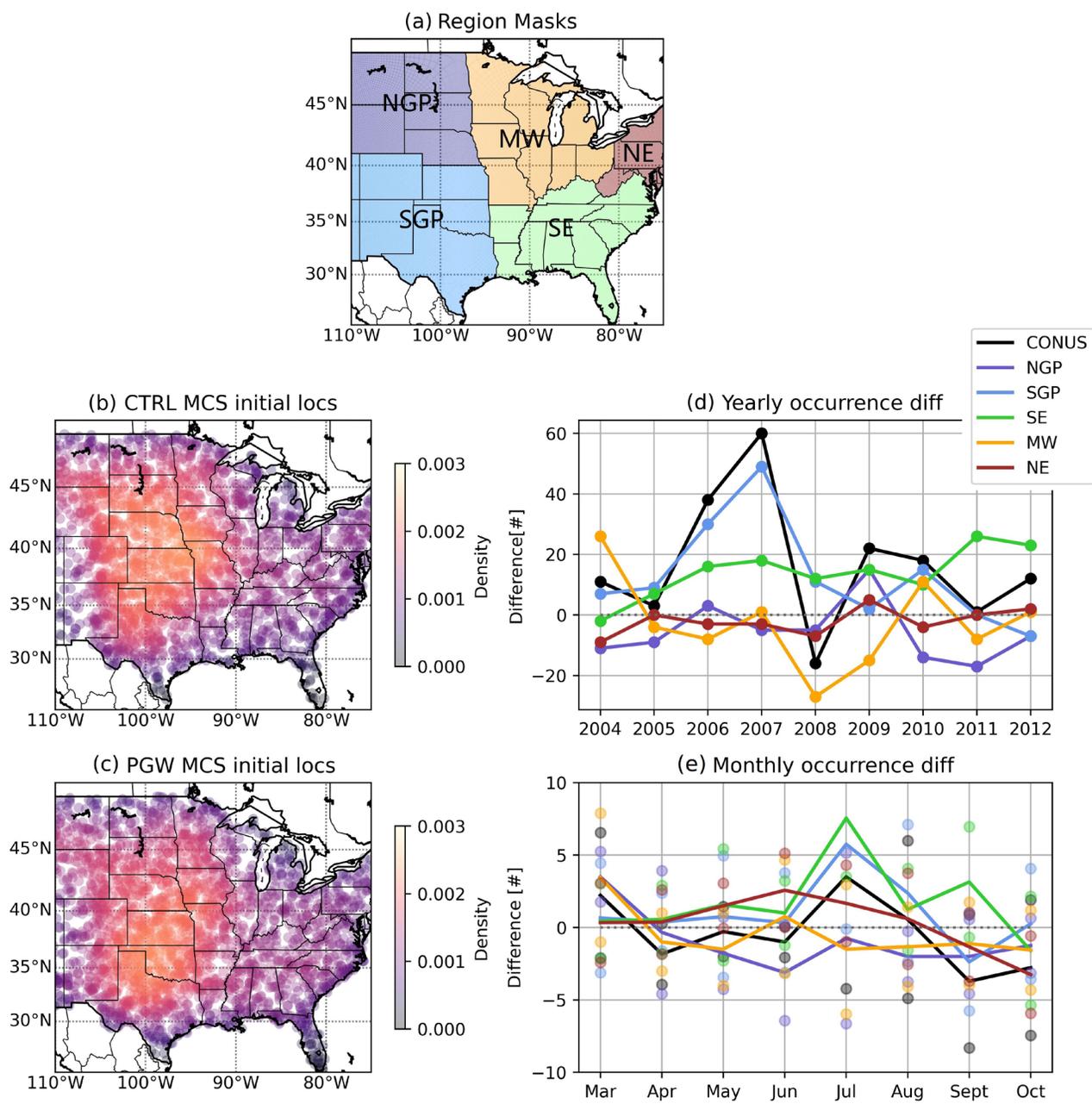

**Figure 3.** Top panel (a) shows the climate regimes defined in this study. Mid-left panel shows the spatial distributions of MCS initial locations color-coded with probability density (scatter density) from (b) CTRL and (c) PGW simulations. Each dot represents the initial location associated with an MCS. Right panel shows the differences (PGW-CTRL) of (d) yearly (March – October) and (e) monthly MCS occurrences between PGW and CTRL runs for different regions: North Great Plains (NGP), Southern Great Plains (SGP), Southeast (SE), Midwest (MW), and Northeast (NE). The dots in the monthly occurrence difference plot represents the 95% confidence interval lower and upper interval values.

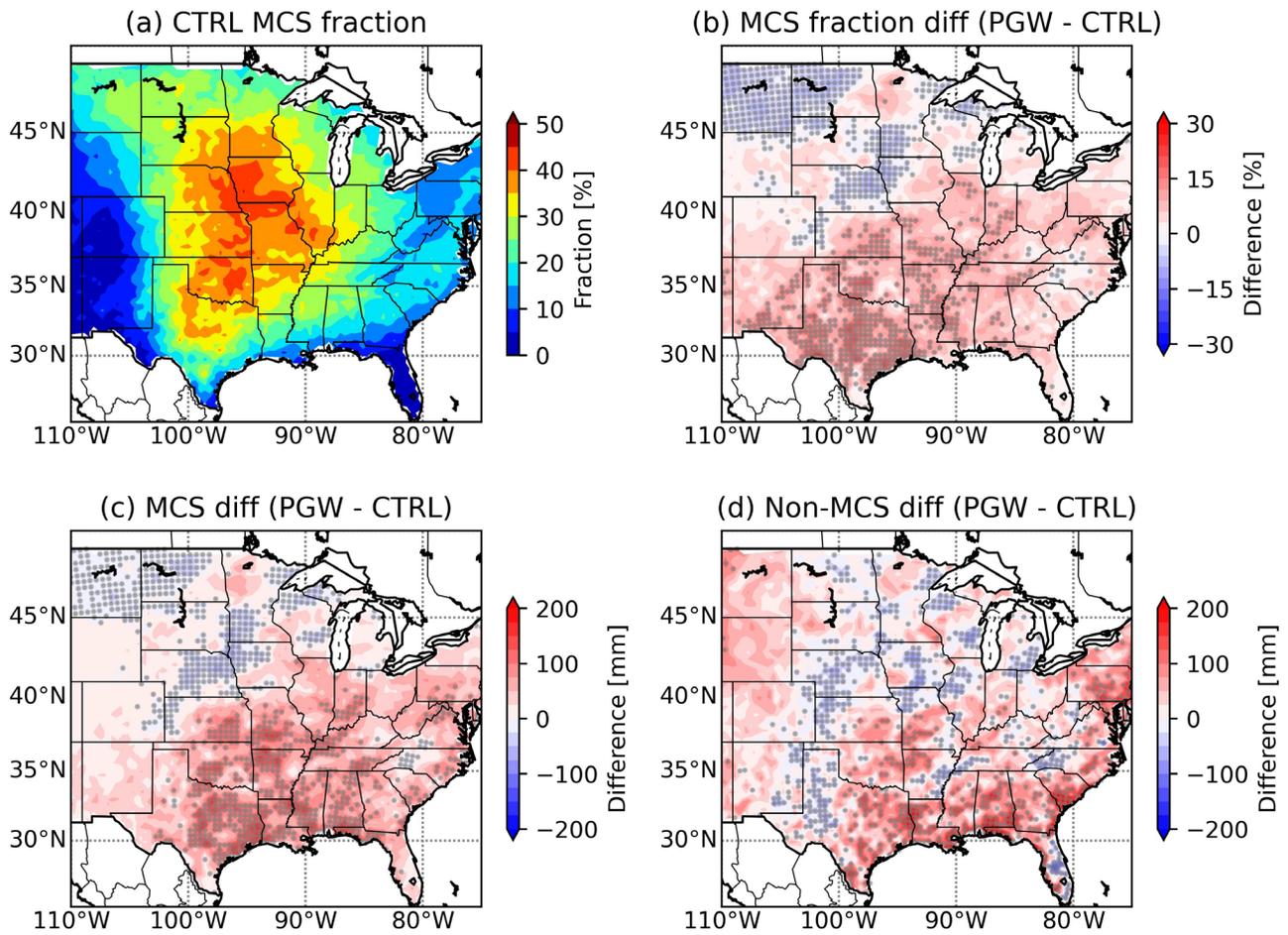

**Figure 4.** Spatial distributions of warm season (March–October) MCS precipitation fraction in (a) CTRL and (b) fraction difference between PGW and CTRL runs, and warm season total precipitation difference between PGW and CTRL runs for (c) MCS and (d) non-MCS events. Grey dots represent the grid values that exceed the two-tailed 95% confidence interval based on the bootstrapping method.

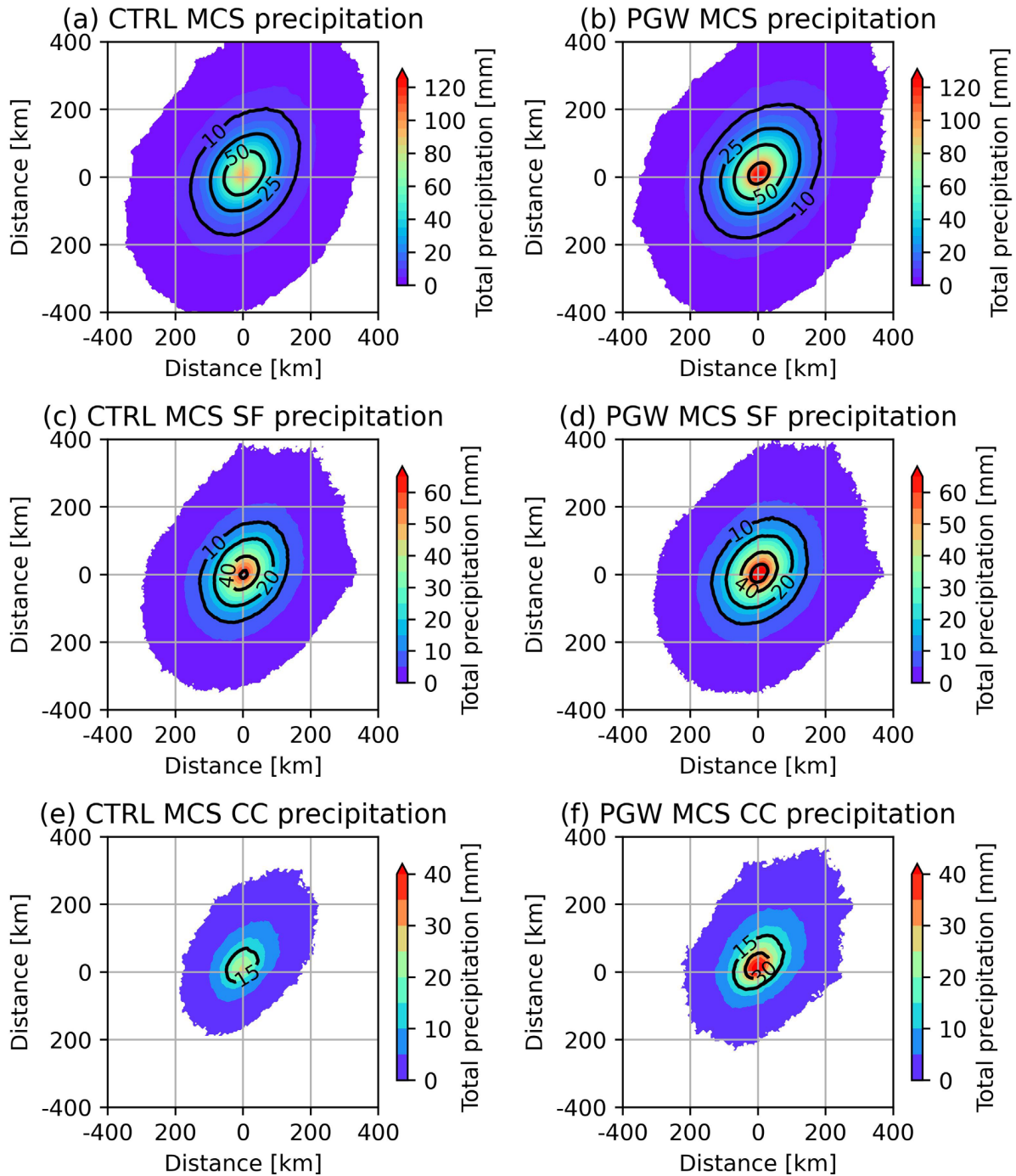

**Figure 5.** Normalized total precipitation (>= 1 mm) from an MCS through its lifetime relative to the storm center (0, 0) from CTRL (left panel) and PGW (right panel) simulations for all (top row), stratiform (mid row) and convective (bottom row) precipitation. All precipitation consists of both stratiform and convective precipitation, as well as all other unclassified precipitation.

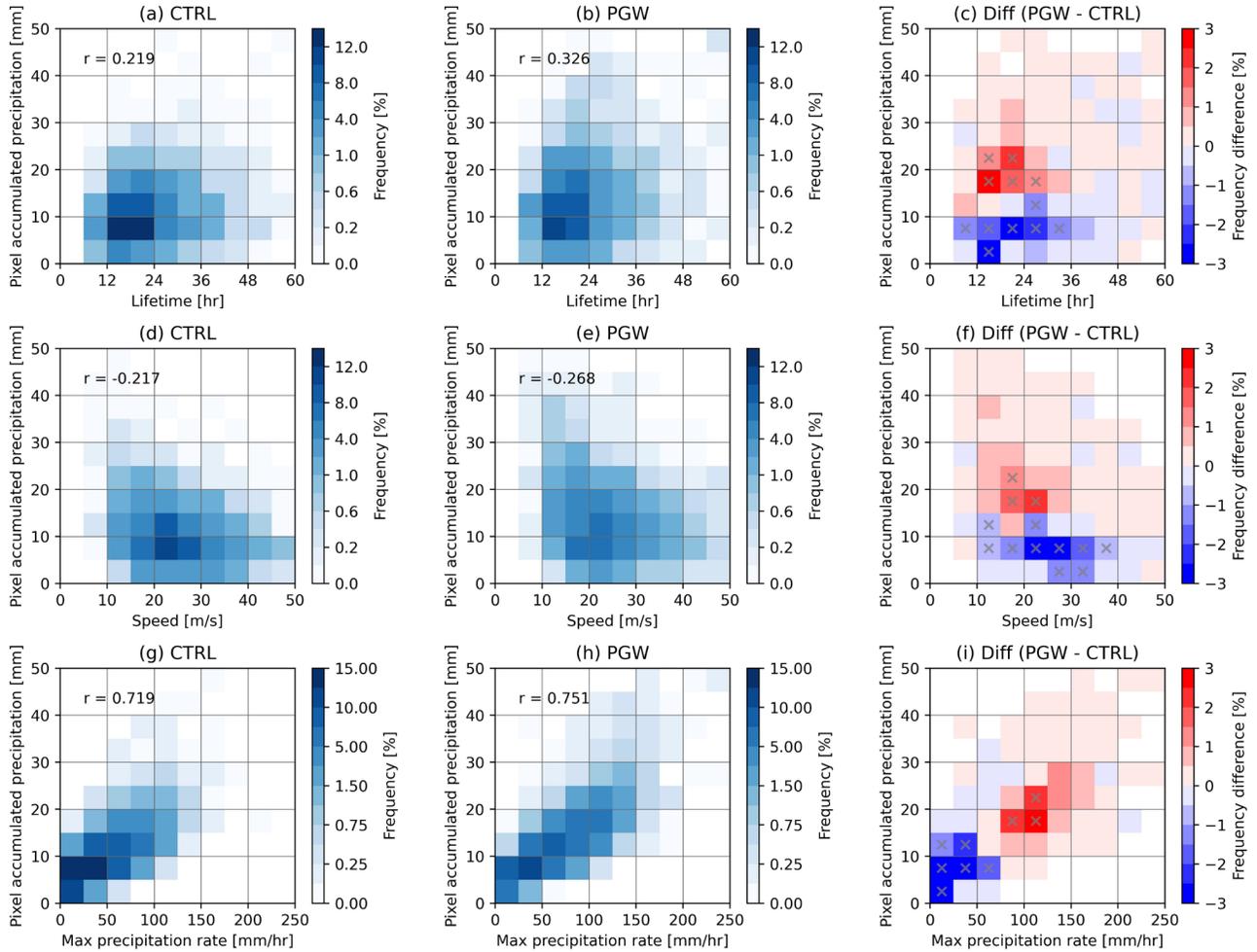

**Figure 6.** Heatmap showing the frequency between (upper) MCS lifespan, (middle) propagation speed, and (lower) maximum precipitation rate against the pixel accumulated precipitation for CTRL, PGW, and their (PGW – CTRL) difference. R value represents the Pearson correlation coefficient between the two paired variables. Grey x mark represents the difference is significant at 95% confidence interval based on the bootstrapping method.

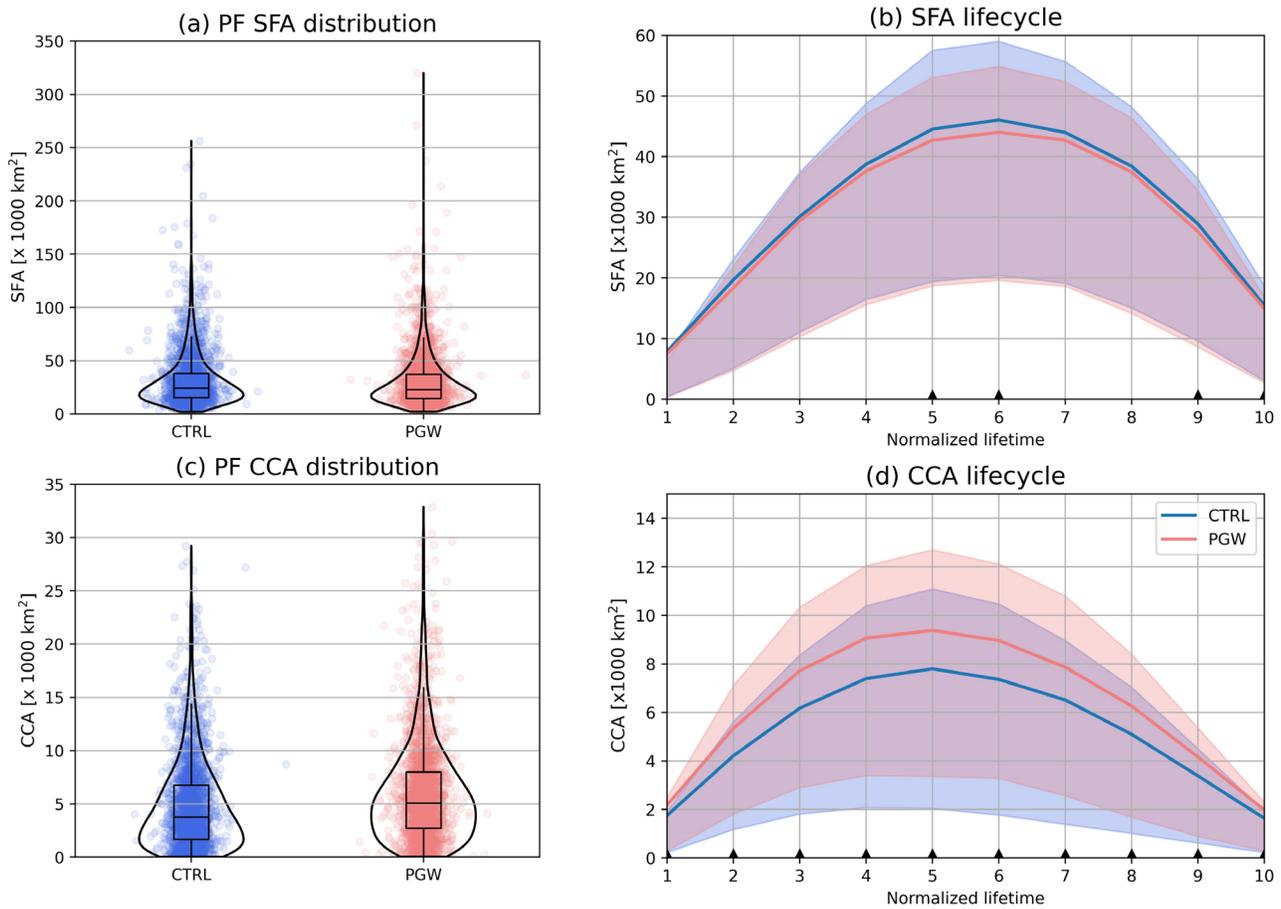

**Figure 7.** Left panel shows the violin plots of the averages of (a) stratiform areal (SFA) size, and (c) convective core areal (CCA) size for CTRL and PGW MCSs. Each dot represents the mean value of SFA or CCA associated with a model-simulated MCS. Right panel shows the (b) SFA and (d) CCA change as a function of MCS's composite lifecycle. The shaded area represents the 25[th] to 75[th] percentile range, the triangle indicates the life stage that the difference between PGW and CTRL is significant at 95% confidence interval based on a Mann-Whitney U test.

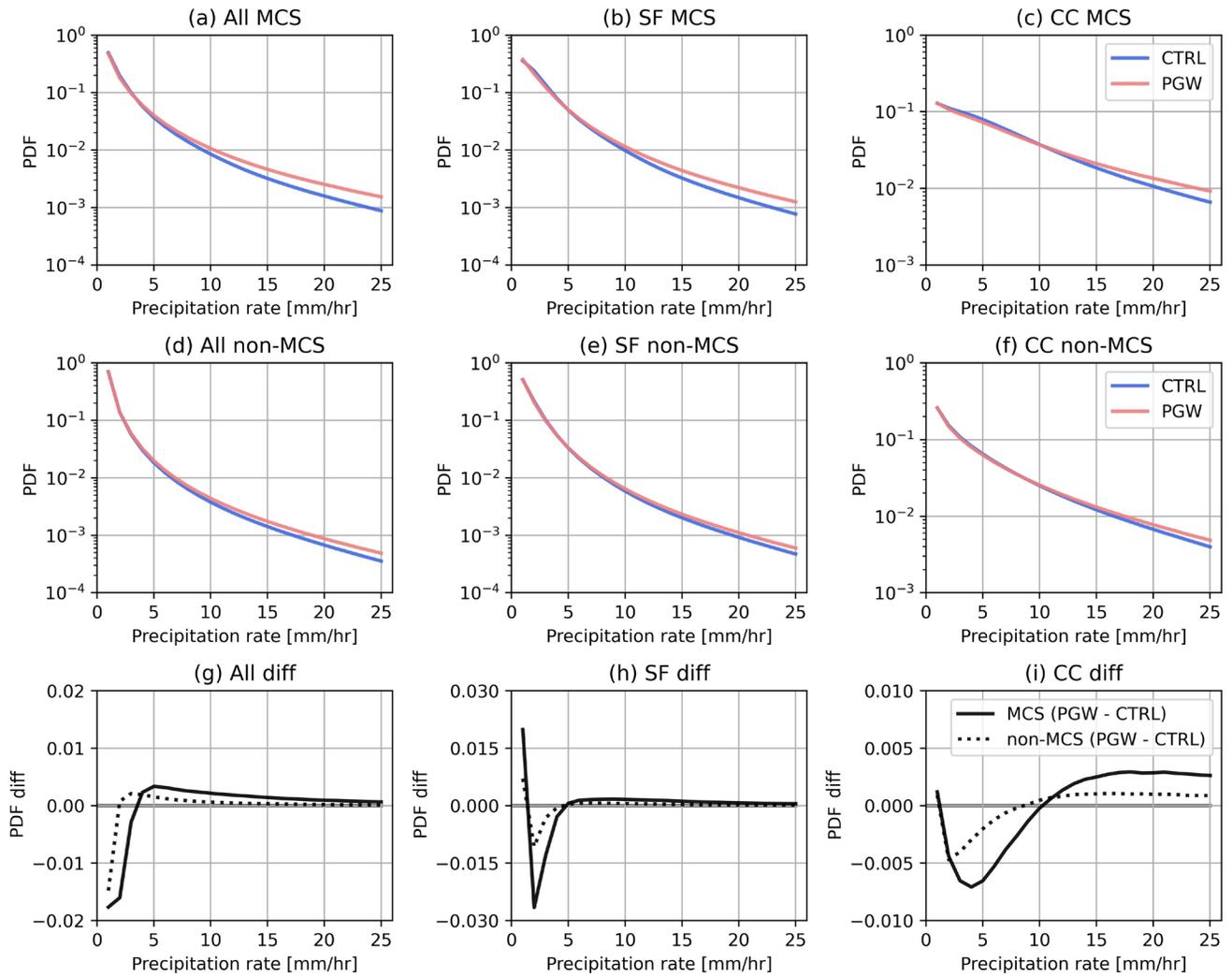

**Figure 8.** Probability density functions of MCS and non-MCS precipitation rate for CTRL and PGW runs, and the difference between two simulations. The comparison is performed for all, and stratiform (SF) and convective core (CC) components of precipitation events. All type consists of both SF and CC pixels, as well as all other precipitating pixels that are unclassified.

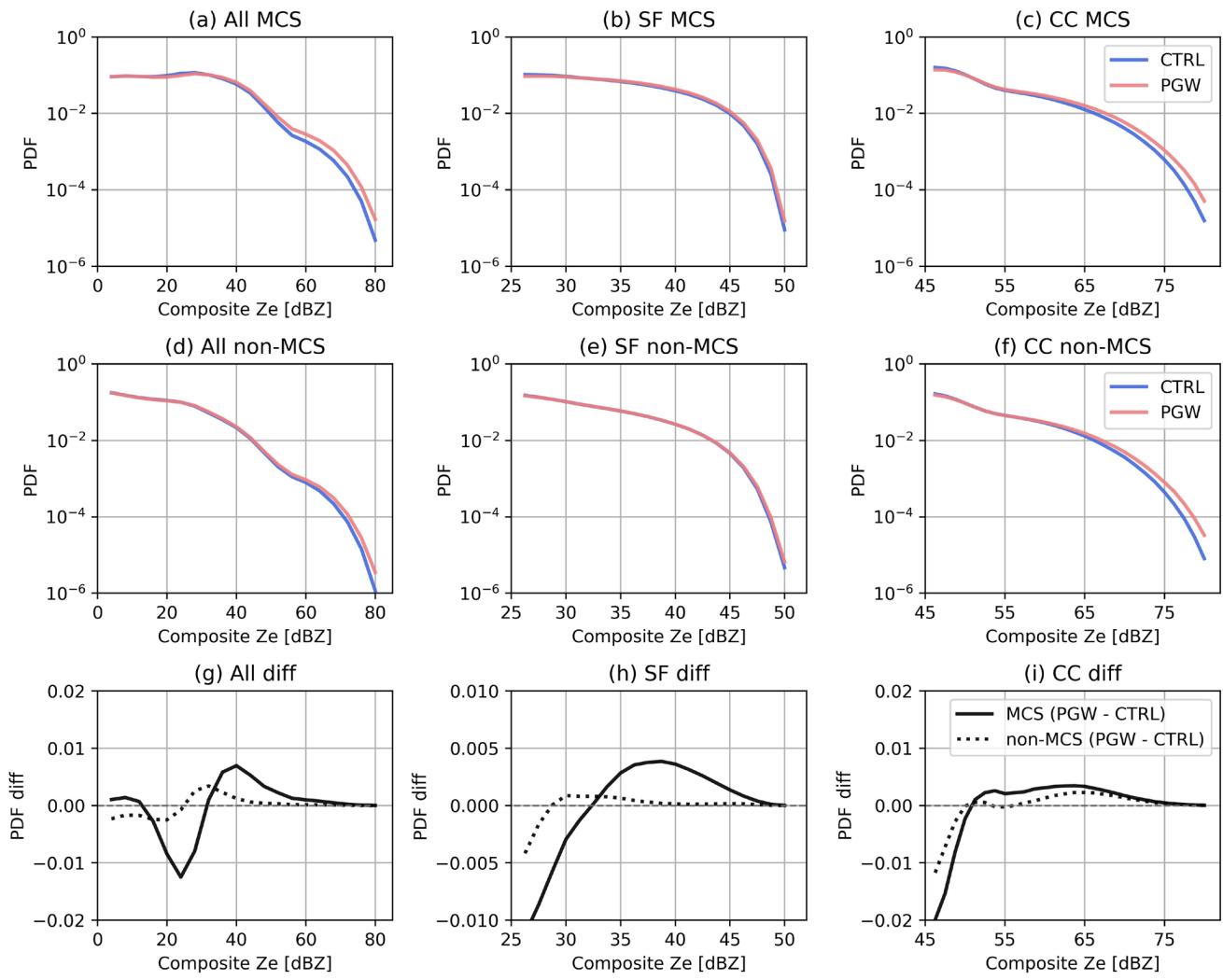

**Figure 9.** Same as in Figure 8, but for composite radar reflectivity.

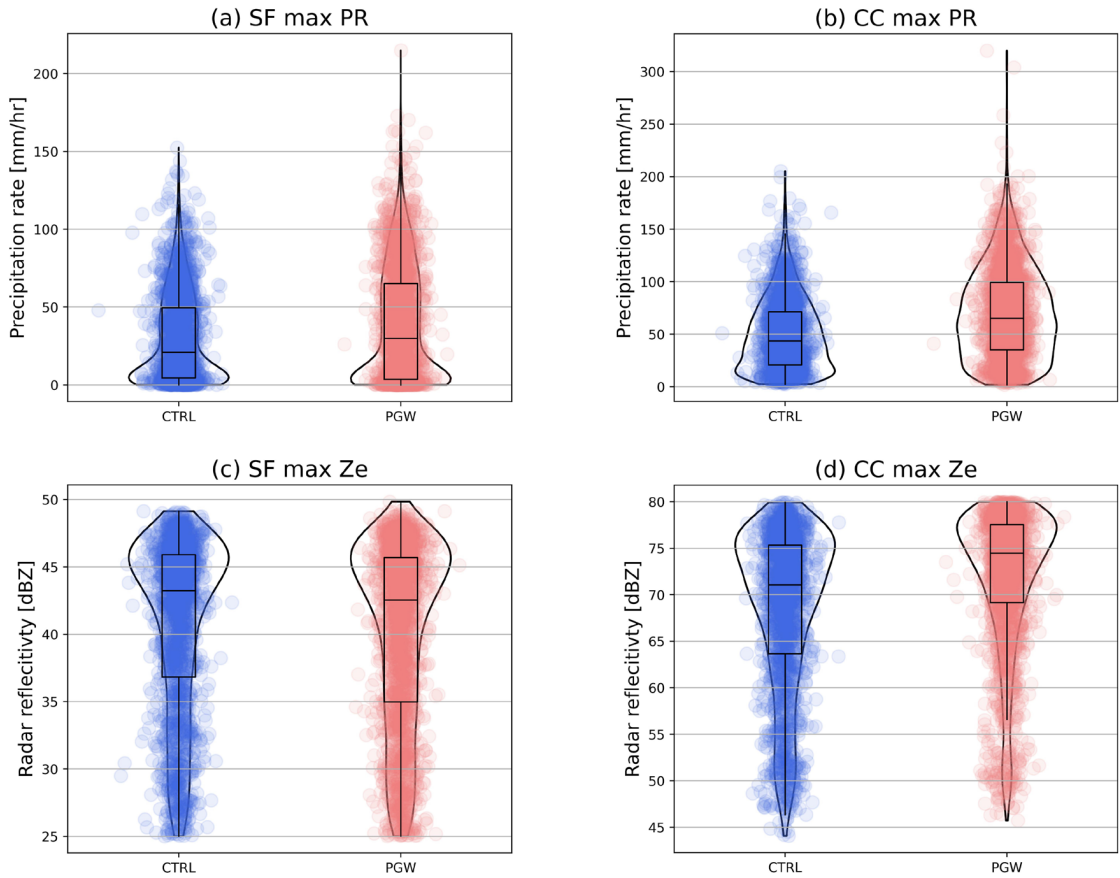

**Figure 10.** Violin plots showing the distributions of maximum precipitation rate and composite radar reflectivity of each MCS in CTRL and PGW for stratiform (SF) and convective core (CC) regions. Each box within the violin shape shows the 25th–75th percentile with the median indicated by the black horizontal line.

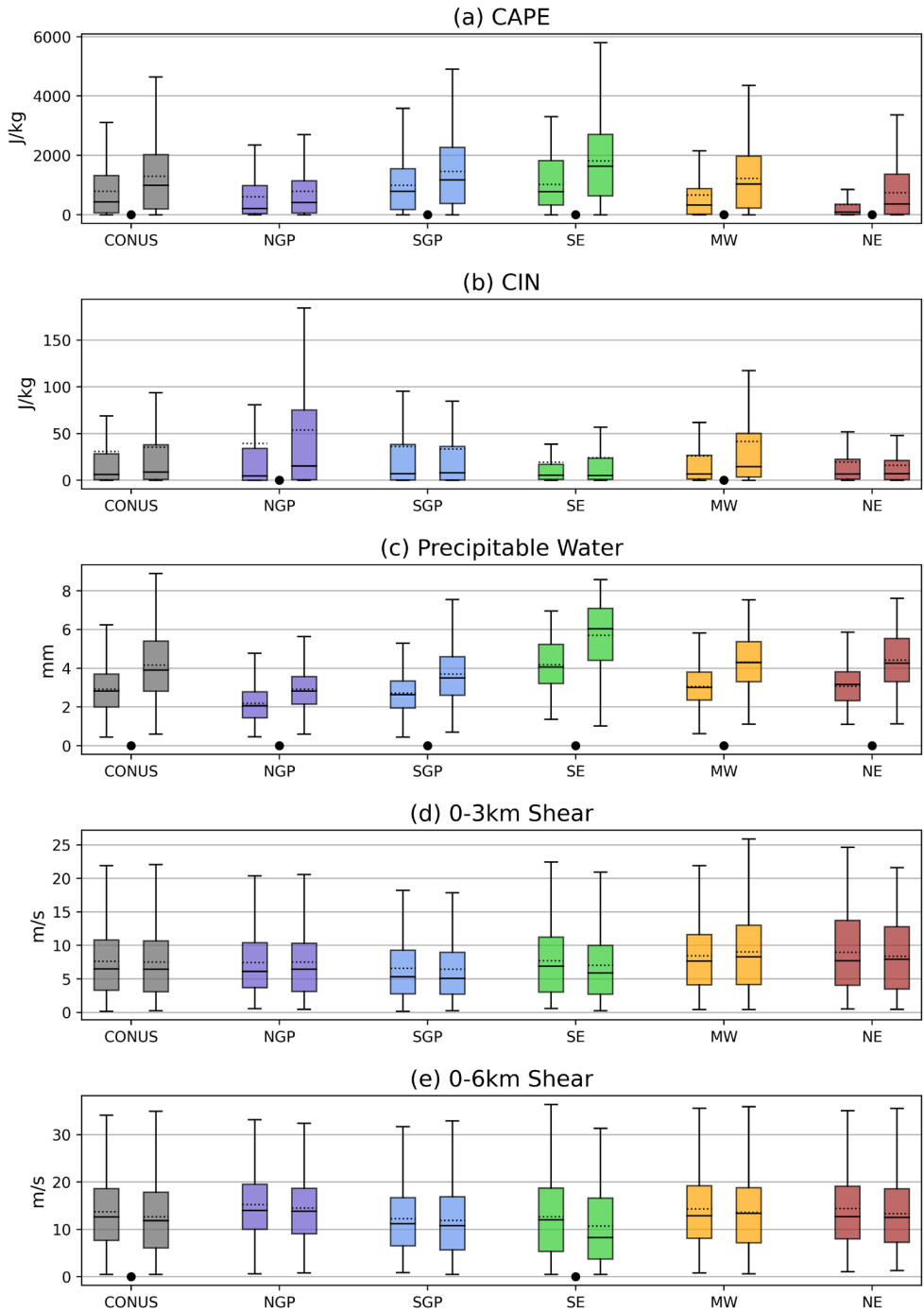

**Figure 11.** Boxplot showing the distributions of area-averaged (a) MUCAPE, (b) MUCIN, (c) preciptiation water, (d) 0-3 km bulk vertical wind shear, and (e) 0-6 km bulk vertical wind shear for MCS pre-initial condition for different climate regions. The left and right boxes of the same color correspond to the values associated with CTRL and PGW MCSs, respectively, for the respective climate region. The box extends from the 25$^{th}$ to the 75$^{th}$ percentile of the data, and the whiskers extend 1.5 times the inter-quartile range from the box. The black line represents the median value, and the black dotted line

represents the mean value. The black dot above the x-axis label name represents the selected variable at specified region, the change from CTRL to PGW is significant at 95% confidence interval based on the Kolmogorov-Smirnov test.

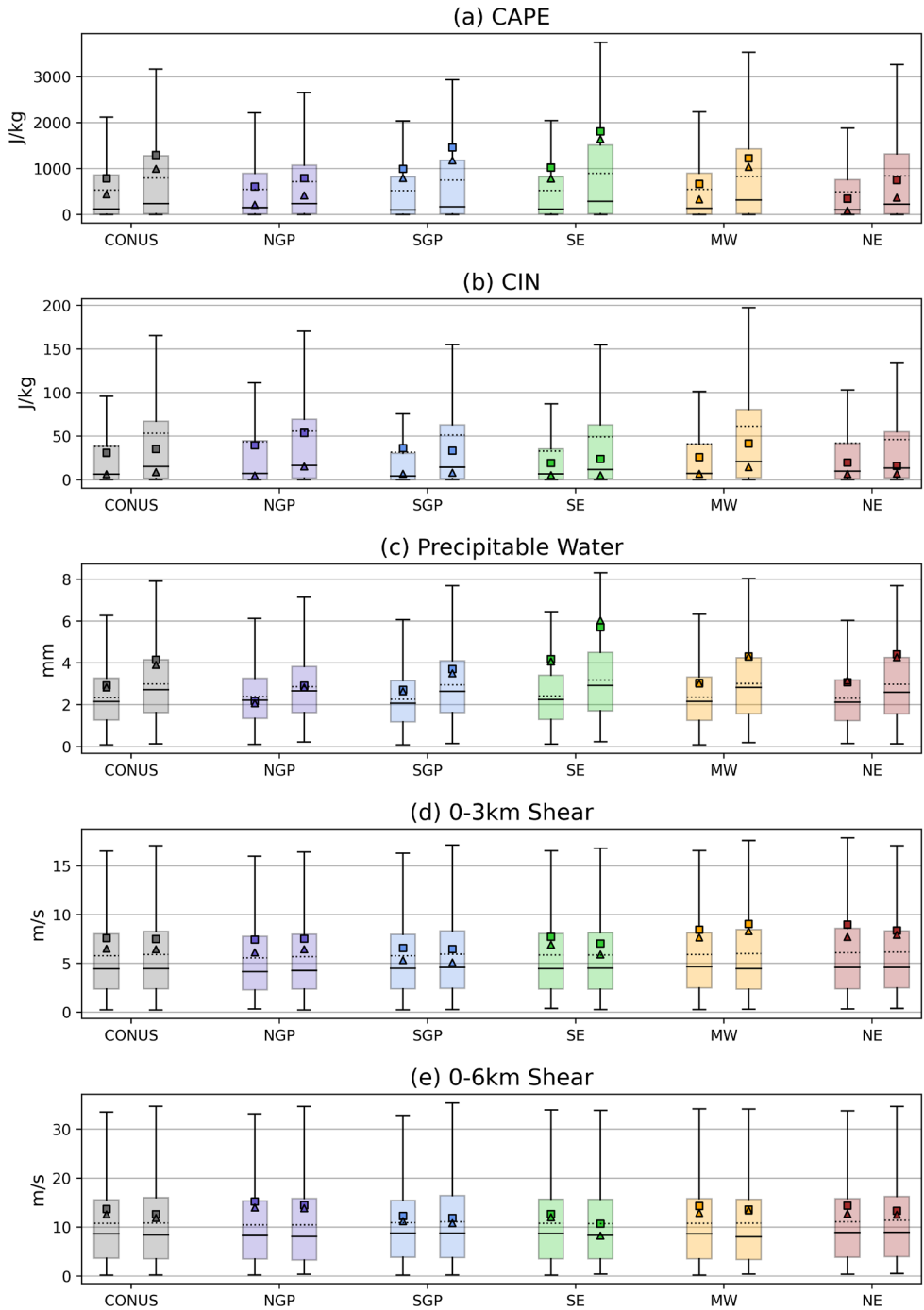

**Figure 12.** Same as in Fig. 11 but for background climatology without the significance test. Triangle and square symbols represents the median and mean of the MCS pre-initial environments in Fig.11, respectively.